**Perspective on Majorana bound-states in hybrid superconductor-semiconductor nanowires.**

*Leo Kouwenhoven*
QuTech and Kavli Institute of NanoScience,
Delft University of Technology, 2600 GA Delft, The Netherlands
*June 25, 2024*

**Introduction.** Topological quantum computing and Majorana bound states were initially theorized between 2000 and 2010. These concepts gradually transitioned to practical implementations during the subsequent decade (2010-2020). Various directions have been investigated with mixed success. With respect to hybrid superconductor-semiconductor devices, great progress has been achieved in the larger area of mesoscopic superconductivity. Firm evidence for a topological phase in hybrid 1D nanowires, however, has not been demonstrated. Now, in the third decade, the lack of definitive topological results prompts a reevaluation. As an active participant, I have witnessed phases of hope, exuberance, and a return to realism and taking a step back. This Perspective provides a personal account of the past decade, my view on the current situation and challenges ahead. I assume the reader is familiar with the subject at the level of, e.g. the review by Prada et al. [1] from which I repeat as little as possible and refer to the extensive reference list for a comprehensive overview. The purpose of this Perspective is to share personal experiences and motivations.

**The hype.** In 2012, during the March meeting of the American Physical Society (APS) in Boston, Nature reported the potential sighting of the mysterious Majorana particle [2]. Just an hour earlier, I had delivered my talk titled "*Signatures of Majorana fermions in superconducting-semiconducting nanowires*" [3]. My concluding slide and words were: "*Have we observed the Majorana particles? I would say a cautious yes*". Within 24 hours numerous news sites, journals etc. repeated the message that Majorana particles had been observed. A hype got started.

The story of the person Ettore Majorana and his particle is a beautiful story full of deep science, personal conflicts of a genius, the mysterious, forever disappearance of the person and the predicted particle that was never found. My APS talk, and the almost live report on the Nature website triggered a viral recount of this story. Great in terms of outreach. Not so great in the expectations it created, many unjustified. I struggled to find a balance between outreach, creating public excitement about our physics, and tempering statements that we 'proved' the observation, instead of the reported 'signatures'. All nuances were gone. To regain some control over the story I agreed to have our daily work filmed. The documentary [4] showed our challenges, mood swings but also the continuous optimism over a six-year period. The dominant feeling was that we had found an opening towards realizing a topological phase and Majorana particles. I believe this optimism was shared by many in the community, both by theorists as well as experimentalists, world-wide. A Perspective from an active theorist is available in Ref. 5. Here I describe my experience as an experimentalist. My account will conclude with the wish that in hindsight my '*cautious yes*' should have included the explicit disclaimer '*that it is way too early for firm conclusions*'.

**The promise.** My first experiments as an undergraduate student in 1986 involved the quantum Hall effect, measuring quantized resistances on two dimensional semiconductors. The quantum Hall effect can be interpreted as a topological phase of matter with a bulk excitation gap and gapless modes at the boundary, known as chiral edge states. Within the realm of the fractional quantum Hall effect more exotic topological phases arise due to strong electron-electron interactions. In 1991 Gregory Moore and Nicholas Read predicted unusual behavior for quasi-particles in a particular fractional state, the 5/2-state [6]. These quasi-particles exhibit non-Abelian exchange-statistics, a stark departure from the familiar Bose-Einstein or Fermi-Dirac statistics observed for all other particles.

In 2003, Alexei Kitaev recognized the potential of non-Abelian quasi-particles for topological quantum computation [7]. Such form of computation is resilient against local noise, making it the holy grail of quantum computing. Despite recent progress on measuring anyons with Abelian statistics [8], non-Abelian quasi-particles remain unobserved in fractional quantum Hall systems. For further reading, I recommend Nicholas Read's review on "*Topological phases and quasiparticle braiding*" [9] and

Nayak et al.'s review on "*Non-Abelian anyons and topological quantum computation*" [10]. I must admit that I struggled to grasp the concepts in Ref. 10, plus I had no access to the high-quality material needed for fractional quantum Hall studies. As a result, I was only following these developments from a remote distance.

**The hybrid platform.** My grasping and engagement changed overnight with the appearance of the 2010-preprints by Lutchyn, Sau and Das Sarma [11] and by Oreg, Refael and von Oppen [12]. Both preprints described how to reach a topological phase with Majorana bound states (MBS) in hybrid combinations of semiconducting (SM) nanowires in contact with standard superconductors (SC). Our group in Delft experimented with such hybrid wires already for some years with high-quality SM nanowires supplied by the group of Erik Bakkers from Eindhoven University of Technology. We had already reported nanowire-based Josephson junctions and SQUIDS and reviewed this field of mesoscopic SC in 2010 [13].

I will refer to Lutchyn et al. [11] and Oreg et al. [12] as the '*nanowire-Majorana proposals*' to indicate the specific 1D approach in solid state. At first glance, these proposals suggested to apply a magnetic field to our earlier devices and a topological phase should arise with a little bit of gate voltage tuning. The proposals seemed to be so close to existing experiments that they did not only arouse our group in Delft but also, among others, in Lund, Illinois and at Weizmann, Purdue and Harvard. The excitement in 2011 is clearly reflected in a News article in Science titled "*Search for Majorana Fermions Nearing Success at Last*" [14].

Also, Frank Wilczek wrote a Perspective on the connection between Majorana particles and SCs in 2009 [15]. Cooper pairing with *p*-wave symmetry (i.e. $p_x+ip_y$) form a topological phase in the bulk of a 2D SC, which comes together with chiral Majorana edge modes at its boundary. In a 1D *p*-wave SC the boundaries are the end points confining the modes to Majorana Bound States (MBS) at zero energy. These zero-energy states can be fractionalized with a half-fermionic state at each end. One such 'half end state' is a MBS and they can be viewed as Majorana particles since they are self-adjoint, i.e. *particle equals anti-particle*. And interestingly, MBS in one or two dimensions obey non-Abelian statistics.

These early theoretical works imposed the idea that Majorana particles and topological phases are two sides of the same coin. It is true that topological SCs do come with some form of Majorana modes. However, Majorana particles can also exist without a topological phase. The self-adjoint definition does not require a topological phase.

Hybrid devices with SMs coupled to SCs can give rise to low-energy states known as Andreev bound states (ABS). In some cases, these ABS are at zero energy where they can mathematically be decomposed in self-adjoint Majorana operators. If the system does not allow for separately addressing these two Majorana parts, then this decomposition has no physical consequences. In this case we simply refer to these states as zero-energy ABSs [16]. But there are also cases where the decomposition does have physical consequences, for example, where one part contributes to tunneling transport while the other does not. Such cases have been found to exist and are denoted as 'quasi-MBS' [17] to indicate their self-adjoint character but without a topological phase in the bulk material. Another non-topological example is the 'poor man's MBS' that can arise in short chains of SM-SC-SM-combinations [18]. Poor man's, non-topological MBS evolve into topological MBS when making the chain longer [19]. Overall, the spectrum of bound states in hybrid materials ranges from trivial ABS via non-topological MBS to topological MBS. This Perspective is intended to give some clarification on this rich regime of physics when combining SMs and SCs. I refer to Ref. 20 for a review on some alternative directions.

**InAs or InSb nanowires.** Given our earlier work on hybrid nanowires, I got invited for a Microsoft Station Q meeting in Santa Barbara in June 2010 where I presented experimental numbers and requirements for realizing the nanowire-Majorana proposals, see Figure 1. The starting point is a SM nanowire with a diameter small enough such that only a few 1D subbands are occupied, ideally just one subband. Next is to contact the nanowire with a SC. This contact needs to be transparent such that Cooperpairs can leak into the SM and proximitize the nanowire with SC correlations.

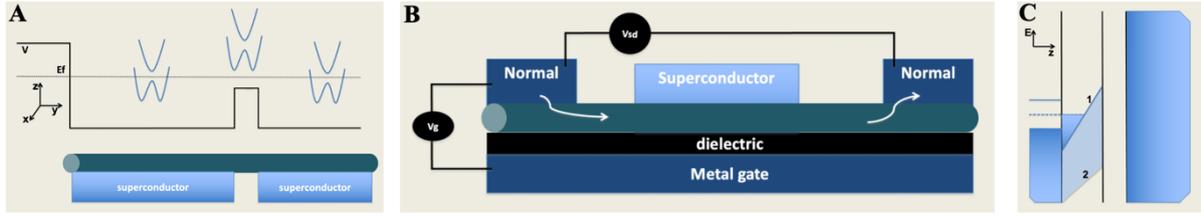

**Figure 1. A**: Schematic device layout of a hybrid 1D wire interrupted by a tunnel barrier. The energy spectrum illustrates the need to line up the Fermi energy in the Zeeman gap of the lowest 1D subband. A magnetic field is applied along the y-direction. The double-minimum in the energy spectrum is a result of Rashba spin-orbit interaction. **B**: Device geometry to perform transport spectroscopy on MBSs. **C**: Energy alignments across the center of the device in **B**. From left to right, schematic Fermi sea with SC gap connected to a nanowire with band bending and accumulation of charge at the interface with the SC [label 1 (2) indicates small (large) amount of charge accumulation]. The right part sketches the Fermi sea of the gate electrode. Pictures from presentation June 2010.

The topological phase requires 'lifting of Fermion doubling', necessary for ending up with just one MBS at each wire end. This can be realized by tuning the electron density with a gate electrode such that the Fermi energy in the nanowire is inside the so-called *Zeeman gap*. A large *g*-factor makes the Zeeman gap, of size $g\mu_B B$, larger and thus easier to tune the Fermi energy inside this gap along the entire nanowire. InAs ($g = -14$) and InSb ($g = -55$) both have large enough *g*-factors, giving a Zeeman gap of order 1 meV for a magnetic field, *B*, of order 1 Tesla.

Potential variations along the nanowire should be less than the Zeeman energy otherwise the Fermi energy moves in and out of the Zeeman gap and consequently not lifting Fermion doubling everywhere. That gives an upper bound ~1 meV to the potential fluctuations arising from the sum of all causes that can induce local potentials (e.g. impurities) or smooth variations (e.g. nanowire tapering). As an example, considering that the variation of confinement energy should be less than ~1 meV, implies that wires with nominally 100 nm diameter should have a tapering less than ~± 2 nm. Or, 2-micron wires should be tapered less than 1:1000. These are back-of-the-envelope estimates, but recent simulations (see Figure 5 below) show these are the correct ballpark numbers.

Another requirement is illustrated in Figure 1C. Connecting a metal to a SM creates band bending for balancing all the electric fields. Depending on the details of the materials (e.g. workfunctions) the band bending can be upwards, downwards or in rare cases, flat. For InAs it is known that the bending is strongly downwards with a pinning of the Fermi energy at the interface of several 100 meV [21]. This pinning is fixed by the interface chemistry and cannot be changed with gate voltages. The lateral confinement, i.e. the slope of the bending, can be tuned with nearby gates. Suppose that under strong confinement with an effective diameter of 20-30 nm the InAs subband spacing is ~10 meV, implying tens of 1D subbands occupied. This metallization [21,22] of InAs nanowires is in my opinion a 'killer' for satisfying the few-subband requirement necessary for a topological phase.

The 'band bending killer' can be resolved by separating the active SM (i.e. InAs) from the SC by an intermediate barrier. Such a complex material stack has been developed recently by the Microsoft team for InAs two-dimensional electron gasses (2DEGs) [23]. 1D InAs nanowires can be grown with a InP shell [24] but such a circular stack has never been optimized properly to have the right barrier that brings the band bending down to less than ~50 meV while still allowing induced superconductivity.

InSb has some favorable materials properties [25]. The *g*-factor, $g = -55$, is about 3 times larger than in InAs. As a result, the Zeeman gap is 3 times larger and consequently the resilience against potential variations increases. In addition, the effective mass in InSb is 3 times lower than in InAs. This increases the 1D subband spacing with a factor 3, making it easier to reach the few-subband regime. InSb does not have a pinned Fermi energy. The amount of band bending at the interface of InSb with Al is unknown. It is believed [25] this bending is an order of magnitude smaller than InAs and thus of order several tens of meV. Together, larger subband spacing (~10 meV for diameter ~100 nm) and considerably less band bending could make it possible to reach the few subband regime.

An additional requirement for realizing a *p*-wave topological SC out of an *s*-wave parent SC is a strong spin-orbit interaction (SOI) in the proximitized SM. It is SOI that can convert *s*- to *p*-wave and the stronger the SOI the better. Both InAs and InSb have reasonably large SOI (Rashba parameter ~10-50 meV·nm) [25].

In 2010 only the group headed by Hongqi Xu in Lund had pioneered quantum transport experiments on InSb nanowires [26]. Such wires are difficult to grow. In MBE, it has not been possible to grow long InSb nanowires with diameters in the range of 50-100 nm and lengths of 5-10 microns. These dimensions are easy to obtain for InAs wires grown in MBE. InSb is better grown with pre-cursor chemicals in vapour-phase chambers (e.g MOVPE). The nanowire-Majorana requirements motivated us to focus on InSb nanowires since MOVPE became available in Erik Bakker's lab. Over the years these wires improved in terms of dimensions (thinner and longer) with lower concentrations of impurities. The current status is ~10-micron long wires with 80-100 nm diameter and measured field-effect mobilities of ~50,000cm$^2$/Vs [27]. Note that mobility measurements in 1D wires must be taken with a grain of salt since assumptions are made for unknown parameters like the capacitance. Nevertheless, despite the uncertainty about the absolute mobility numbers, nanowires have shown a consistent improvement, and they may well be in the range of satisfying the nanowire-Majorana requirements for materials properties, wire dimensions and mobility [28].

**The SC: NbTiN or Al**
Next is the choice of SCs. We initially chose for NbTiN since (a) it was available in Delft in the group headed by Teun Klapwijk, (b) it has a large superconducting gap of about 3 meV [25], and (c) it can sustain large magnetic fields. NbTiN is a dirty SC with a very short coherence length of just a few nm. How this affects topological properties was and I believe is still unknown.

Wolfgang Pauli's quote '*God made the bulk; surfaces were invented by the devil*' also applies to SM-SC interfaces where the '*devil is in the details*'. To understand the experimental route towards this interface, one needs to realize that the wires are grown in Eindhoven, then shipped to Delft while being exposed to ambient conditions, then mounted in a metal deposition chamber. We pump out the gasses, remove oxides using an Argon milling etch and then deposit NbTiN by sputtering. Along this route, pristine wires get very dirty and need to be cleaned to atomic levels before covering them with the SC. In principle this is not impossible but very hard with Argon milling which removes oxides by bombarding the surface and kick out the top atoms by force. Our group has spent an incredible amount of time performing the cleaning step as gentle as possible, minimizing damage on the InSb nanowire [29]. I believe we went from bad (leaving In droplets on the nanowire surface) to pretty good (although I don't think we ever attained atomically perfect interfaces); see Figure 2. Since it is not possible to image the curved interface after NbTiN deposition, the final structural quality is unknown and can only be inferred indirectly from transport characteristics.

Devices with rough interfaces yield soft induced SC gaps, i.e. gaps with a continuum of subgap states [30]. Nevertheless, such devices led to our first observation of zero-bias peaks (ZBPs) in 2012 [31]. Later, improved devices with cleaner interfaces (such as lower Left panel in Figure 2) resulted in a much harder SC gap and ZBPs persistent over extended regions in parameter space. Figure 2 taken from Ref. 32 shows one of our best results on InSb-NbTiN. The ZBP extends over more than a meV in Zeeman energy, which is more than 20 times the linewidth of the ZBP. Observations like this, data taken in 2016, made us conclude that these ZBPs did not originate from accidental crossings of Andreev bound states [33]. Also, such robust ZBPs did not seem to be expected for disordered wires and phenomena like anti-weak localization [34]. We felt in 2018 that 'all known alternative explanations, other than a Majorana explanation' were excluded by these results [32]. We return in the discussion section to the question if, at the time, we could have been aware of the possibility of "unknown explanations". A relevant question since we could have emphasized more that our hybrids were largely a "black box" with many unknowns, including the electrostatic potential landscape with all the sources for disorder and the nature of induced superconductivity from a dirty SC like NbTiN. I have chosen to reproduce Figure 2 here since the robustness of the ZBP in this data still intrigues me.

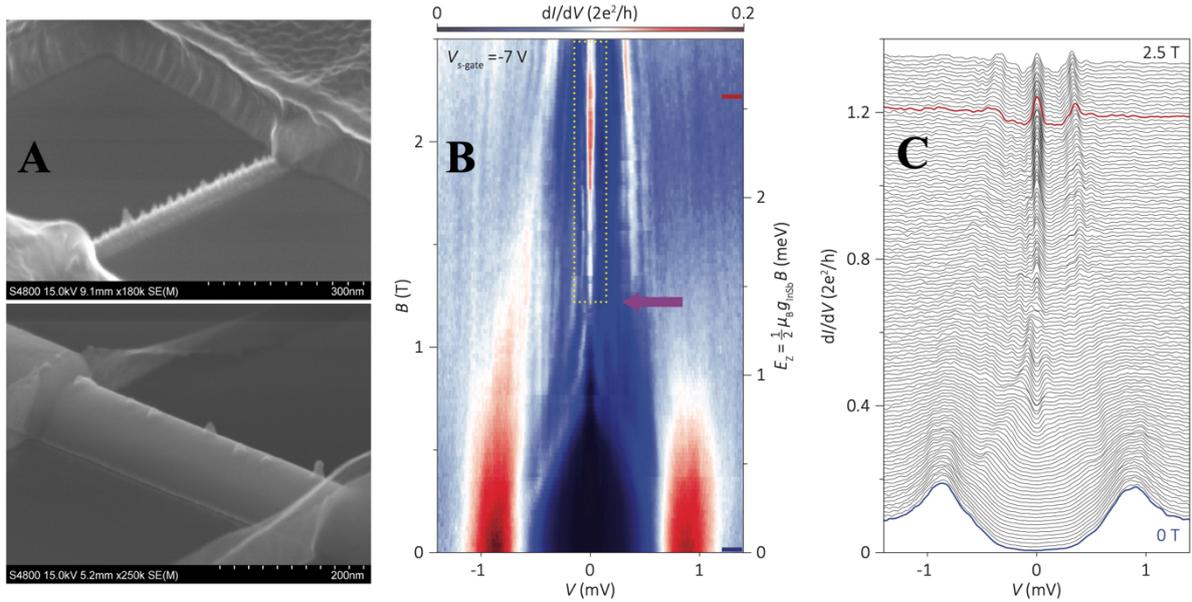

Figure 2. Top Left: Too much Argon milling of InSb nanowires can leave behind In droplets. Bottom Left: Gentle milling leaves a clean surface but still with an occasional droplet. Right panels show exemplary ZBP data on InSb-NbTiN in color scale and as line-cuts. From Ref. 32.

A breakthrough development occurred in 2015 with Krogstrup et al. [35] reporting epitaxial growth of Al on InAs nanowires. The Al was deposited on the pristine InAs wires *in-situ* in the MBE vacuum chamber and thus they could avoid the etching cleaning step. They obtained an atomically clean interface with epitaxial Al on the InAs crystal. The transport characteristics, measured in the group of Charles Marcus in Copenhagen, improved enormously with beautiful ZBPs [36] and Coulomb islands with zero modes [37]. At the time, these results seemed consistent *only* with a topological MBS interpretation.

Al cannot be grown in-situ directly on InSb since vapour-chambers are not compatible with the vacuum chamber for Al growth. Nevertheless, inspired by the success of Ref. 35 the materials groups of Erik Bakkers and Chris Palmstrom (UC Santa Barbara) found a work-around. The InSb wires are sent to Santa Barbara where the oxide is removed by hydrogen-etching. In contrast to Ar milling, H-etching is a gentle chemical process where H radicals bind to Oxygen atoms and then disappear from the surface into the vacuum chamber. After cleaning with near-atomic precision, the wires are moved within the same vacuum-space to the Al chamber. This process also yields something close to epitaxial growth of Al on InSb [38].

When my group members saw the TEM pictures of the epitaxial Al on InSb there was an immediate drive to make devices. The first results, however, were disappointing, no induced superconductivity. It turned out that during fabrication the substrate temperature increased, inducing a diffusion process of Al into the nanowire forming insulating AlSb at the interface. The thermal budget available for fabrication that keeps the epitaxial InSb-Al interface intact is very low (i.e. room temperature), much lower than for InAs-Al. This makes InAs-Al the much-preferred combination over InSb-Al with respect to *ease of fabrication*.

Nevertheless, we continued with InSb-Al and only used room-temperature processing, e.g. no resist baking. This can be done but at the expense of lower quality resists and dielectrics. As a result, the yield of useful devices was low (~10%) and even the good devices still suffered from charge switches. Still some devices, with a yield <5%, showed transport properties much better than we had ever seen on InSb-NbTiN. We found hard superconducting gaps, enhanced Andreev transport, Coulomb islands with zero modes and ZBPs with maxima as large as $2e^2/h$ and even above this quantized value [39].

**Intermezzo Retracted Nature Articles**

The story of the ZBPs at $2e^2/h$ does not have a happy ending. At the time (2017), the quantization of the zero-bias conductance at $2e^2/h$ was important evidence for MBSs. In our Nature publication (2018) titled "Quantized Majorana conductance" we claimed the observation of this proof. Initiated by Sergey Frolov and Vincent Mourik, a re-analysis of the data in 2020, however, showed serious shortcomings and we retracted our publication [**A**]. Mistakes were also made in characterization measurements of the growth paper, also published in Nature, and also retracted [**B**]. The two retractions had technical and ethical aspects. On the technical side, one type of mistake consisted of unmentioned corrections for charge switches. Data sets were cleaned up by deleting irregularities suspected to origin from charge switches. This results in high-quality looking data although the underlying raw data contained a serious number of switches. Another serious technical error was a miscalibration of the conductance value by almost 10%. The recalibrated data moved some data even above the quantized value, thereby falsifying the claim of a quantized Majorana conductance. We corrected these technical errors in a rewritten manuscript [**C**] together with extended figures showing more data than in the retracted version. The "Rewrite" reported large ZBPs with heights of order $2e^2/h$ instead of quantized peaks and no distinction could be made between enhanced Andreev transport and a Majorana conductance.

For the ethical aspect, the retraction was followed with an integrity investigation by independent experts [**D**], by the committee of scientific integrity at Delft University of Technology as well as at the Dutch national committee of scientific integrity [**E**]. No evidence was found for intentional mistakes or scientific misconduct [**F**]. Nevertheless, the experts believed that the authors suffered from confirmation bias in the sense that they were selectively focusing on observations of quantized conductance and ignoring evidence against this. Indeed, the larger than published data set contained ample evidence against the claim of a quantized Majorana conductance. In conclusion, these larger data sets should have been discussed explicitly at the time of publication. This should have been accompanied with a more careful and balanced discussion of the interpretation of these large ZBPs, as was corrected in [**C**]. For all these errors, oversights, and misrepresentation, I owe the community an apology. Furthermore, I believe the entire retraction process also deserves, in due time, an extended Perspective on its own.

**Links**
**A.** Retracted Zhang et al.: https://www.nature.com/articles/nature26142.
Corresponding Retraction Note: https://www.nature.com/articles/s41586-021-03373-x
Data Repository: https://zenodo.org/records/4545577
**B.** Retracted Gazibegovic et al.: https://www.nature.com/articles/nature23468
Corresponding Retraction Note: https://www.nature.com/articles/s41586-022-04704-2
Data Repository: https://zenodo.org/records/5025868
**C.** Rewrite of Zhang et al.: https://arxiv.org/abs/2101.11456
**D.** Expert report integrity investigation: https://zenodo.org/records/4545812
**E.** Conclusions integrity investigation by Dutch national committee (LOWI): https://lowi.nl/advies-2022-03-en-04/
**F.** Conclusions integrity investigation by Delft University of Technology:
https://www.universiteitenvannederland.nl/files/publications/2022%20TUD%20Onzorgvuldig%20en%20verwijtbaar%20on zorgvuldig%20handelen%20wat%20betreft%20de%20selectie%20van%20data%20maar%20geen%20schending%20van%2 0de%20wetenschappelijke%20integriteit%20-%20ongegrond.pdf

The low yield and charge switches made it clear that this path of room-temperature fabrication had no long-term future. We knew of one solution, but it would involve years of re-developing a new fabrication scheme where all processing was done *before* creating the delicate InSb-Al interface. A similar inverse fabrication scheme [40] had been hugely successful for our work on carbon nanotubes [41]. Moreover, instead of a blanket Al deposition with subsequent etching, we opted for *selective deposition*. We fabricated 3D objects on the substrate and placed the nanowire in their shadows. Using these shadows by carefully chosen incident angles of the Al flux, we realized half-covered nanowires with a desired length for the hybrid sections anywhere between 200 nm to 8 microns. Figures 3 and 4 show some examples of devices made with shadow walls. Gate electrodes are already buried inside the substrate with a covering dielectric grown at an optimized temperature (~ several hundred degrees Celsius); since the new inverse fabrication is not limited by a thermal budget. This mitigated the issue of charge switches and consequently yielding much more reproducible transport characteristics.

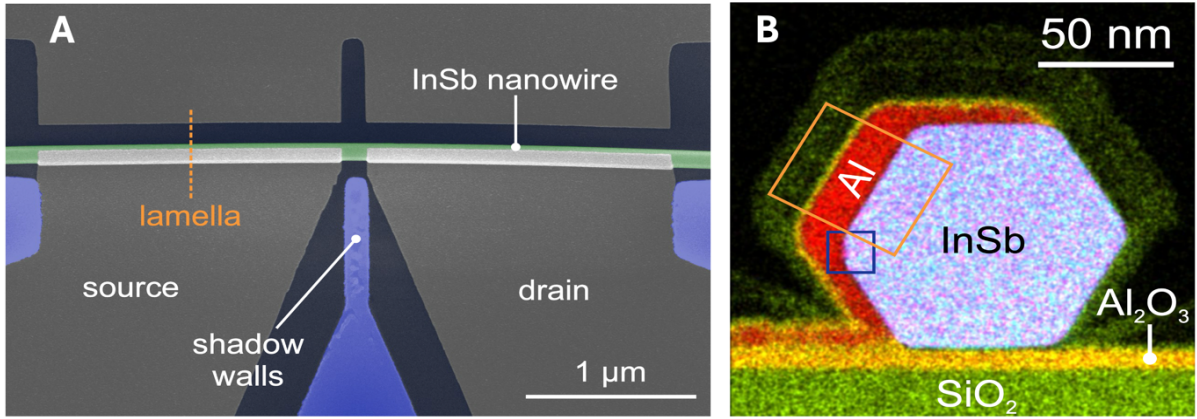

Figure 3. **A.** Smart wall device consisting of two hybrid leads and a Josephson junction in the middle (same device geometry as in Figure 1A). The dashed line indicated with 'lamella' points at the position where the cross-sectional image and composition is taken, shown in **B**. From Ref. 42.

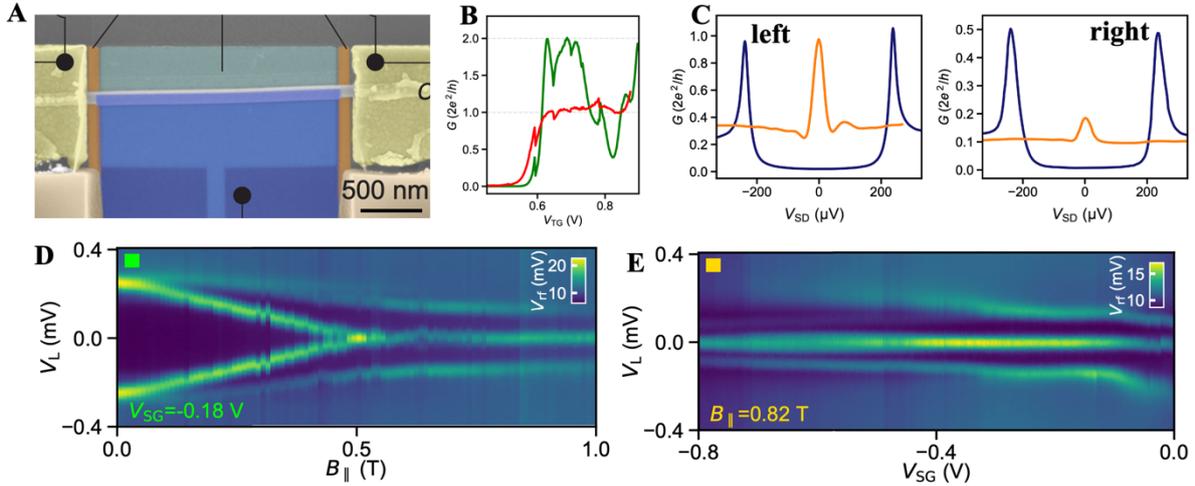

Figure 4. **A.** Smart-wall device with two normal contacts for spectroscopy at the two ends, separated from the hybrid with short sections controlled by tunnel gate voltages (orange); same device geometry as in Figure 1B. The nanowire is half covered with Al (blue; similar cross section as in Figure 3) which is connected to ground. The length of the hybrid section is ~2 micron. **B.** Conductance quantization of high quality for such a hybrid system. Red curve is for out-of-gap normal transport (ideally quantized at $2e^2/h$) and green curve for in-gap Andreev transport (ideally quantized at $4e^2/h$). Data taken for $B = 0$. **C.** $B = 0$ (blue) and finite $B$ (orange) spectroscopy traces, illustrating a hard gap at $B = 0$ and ZBPs at non-zero $B$. Two panels show simultaneously taken data from opposite ends of the hybrid [42]. **D and E.** Bottom panels show two exemplary measurements of zero-bias peaks that remain at zero bias for magnetic field between 0.5 and 1 T and in gate voltage between -0.8 and 0 V. From Ref. 43.

I show the data in Figure 4 as examples of our best results after ten years of improving device quality. The ZBPs are large and stick to zero energy over large ranges in magnetic field and gate voltage, many times larger than their linewidth. We sometimes find ZBPs at both ends simultaneously after an extensive search (e.g. Fig. 4C). The yield of devices exceeds 80% and charge switches rarely happen. Overall, the robustness and quality of the ZBPs have improved enormously since our 2012 results.

The interpretation of ZBPs, however, also evolved over the years. We no longer think that the ZBPs in Figure 4 reflect two MBSs at the two wire ends, separated by an uninterrupted topological phase. We now realize that the nanowire-Majorana model really is just a minimal model for a theorized hybrid wire. In hindsight, this minimal model created a too simplistic mindset that made us take too large steps instead of seriously investigating the basic issues that I will discuss in the next paragraphs.

A final note in this section on the parent SC. NbTiN is a dirty SC with granular morphology and, we now know, unsuited to serve as the parent SC for realizing an induced topological phase in the nanowire-Majorana approach. This statement is further substantiated by all our attempts to create SC

Coulomb islands with NbTiN. We only obtained 1$e$-charge periodicity, indicating the abundant presence of poisoning quasi-particles. In fact, I am not aware of any 2$e$-charge periodicity from Nb-based Coulomb islands. The Al results by Krogstrup et al. [35] suggested the necessity of an epitaxial SC-SM interface. Our results, however, with exemplary data in Figures 3 and 4, are obtained in a non-MBE deposition system. The Al is of high quality with quite homogeneous thickness along the hybrid. The morphology is mostly granular with here and there some local epitaxial relation between SM and SC. By no means, however, is our Al epitaxial and single domain over long length scales, e.g. ~100 nm or more. It is now clear that an epitaxial SM-SC interface is not necessary for a hard induced gap, enhanced Andreev transport, robust ZBPs or hybrid Coulomb islands with 2$e$-charge periodicity. Non-epitaxial Al can also give these results [42]. The critical distinction between e.g. NbTiN and Al is not epitaxy, but instead because Al is a very special SC. Nearly all SCs have a diminishing gap when reducing film thickness or adding disorder. Al stands out with an increasing gap when film thickness is reduced below ~10 nm, reaching roughly twice the bulk gap value for a thickness of ~2 nm [44]. This corroborates with the fact that (non-hybrid) SC qubits are all made with Al/AlOx/Al Josephson junctions, without exception. The microscopic reason for a larger gap in Al in thin films is unknown, as far as I know. One could argue that thin or dirty films have a modified phonon spectrum but how exactly this yields a larger gap is yet unknown and would certainly deserve a deep dive study. Fortunately, this mysterious property turns out to be extremely positive for the realization of both hybrid and qubit devices, and so we all make use of it.

**The bulk I: proper density and coupling strength.** If we consider a 1D hybrid completely free of disorder, there still are several requirements to satisfy the conditions for a topological phase. The Fermi energy needs to be inside the Zeeman gap. This requires tuning of the gate voltage controlling the electron density, since the Zeeman gap (~1 meV) is an order of magnitude smaller than the 1D subband spacing. While knowing the size of the Zeeman gap, unfortunately, we do not know where in gate voltage the Fermi energy is inside this gap and thus tuning can only be done by searching for a specific outcome.

In addition, the SM-SC coupling needs to have an appropriate strength. Very negative gate voltages push the SM electrons against the interface such that they become strongly hybridized and obtain (ABS) energies close to the parent SC gap. In this case of strong coupling, electrons lose their SM properties and adopt the $g$-factor and SOI strength of the parent SC. In this limit the magnetic field needed to create a Zeeman gap is at the same time destructive for the parent SC.

In the opposite limit of very weak coupling, electrons do retain the required $g$-factor and SOI, but the induced gap is very small. Effectively the electronic states become weakly proximitized and their 'gap energy' may be dominated by finite size effects. The results from Albrecht et al. [37] reporting zero-modes with exponential splitting behavior seem to be in this regime of weak coupling where the splitting is due to finite-length effects instead of a topological origin. This finite-size splitting instead of topological scaling is consistent with follow-up experiments as well as theoretical modeling [43,45,46].

The optimized coupling strength is in between these two extreme limits and roughly at a value where the induced gap at $B = 0$ is about half the strength of the parent Al gap [23,45]. In this regime, the electronic states retain SOI, and although their $g$-factor is modified, also roughly half the bulk value, it can still be much larger than the $g$-factor of the parent SC. According to the nanowire-Majorana minimal model for the case of a SOI energy much smaller than the induced gap, the maximum topological gap can reach about half the $B = 0$ induced gap. So, relative to the parent SC one loses a factor of four. To put in some numbers, an Al gap of ~ 200 μeV results in a maximum topological gap of ~50 μeV.

Nanowire hybrids have only one gate to satisfy both the Zeeman gap tuning as well as the coupling tuning. There is no reason that both requirements are satisfied at the same gate voltage. There are attempts to resolve this issue by growing thin tunnel barriers as a shell around the nanowire with thicknesses optimized for the appropriate coupling strength, but so far not satisfying the nanowire-Majorana requirements [47]. The absence of a good barrier for InSb nanowires made my group decide in 2022 to stop our research on hybrid InSb-Al materials for the purpose of realizing the nanowire-Majorana proposal. A last publication reporting electrostatic control of the induced gap [48] also made it clear that a simultaneous appropriate coupling strength is unattainable in a controlled manner in this material stack.

**The bulk II: disorder from impurities.** Next, we consider random disorder due to impurities. We assume that everything else is perfect, including an optimized induced gap and the ability to tune into the Zeeman gap. We also ignore long-range inhomogeneities such as tapering.

Short-length scale disorder can come from numerous sources. First, imperfections in the semiconducting material like vacancies, impurities and surface oxides [27]. Nanowires do have the advantage to grow free of strain. In contrast, 2DEGs grown on insulating substrates have additional sources of disorder due to strain-induced misfits leading to all kinds of dislocations with crosshatches as just one example which are particularly destructive when defining 1D wires [49]. Substrate-free growth is a serious advantage of nanowires over 2DEGs.

Pristine nanowires are placed on an insulating substrate and imperfections in the substrate have a long-range electrostatic effect. Nanowires are often in contact with a dielectric which always have some finite density of trapped charges. Moreover, the interface with the SC is most likely not perfect and contains atomic scale variations. The SC also contains a large amount of disorder from both surface oxidation and grain structure. It is believed that for weak and intermediate coupling, disorder in the SC is not induced into the SM [50,51]. Last, but not least, fabrication introduces additional disorder, for instance, from all kinds of chemical residues. Also gate electrodes have corrugated edges of nanometer scale, e.g. set by the metal grain structure, and as a result the induced gate potential is not perfectly homogeneous.

All these short-length scale causes for disorder add up to a resulting, self-consistent potential in the nanowire. Direct measurements of this potential landscape do not exist. Maybe there is a huge amount of disorder or maybe the nearby SC efficiently screens potential variations; this is largely unknown. Microsoft has extracted one number from subgap transport measurements yielding a localization length exceeding 1 micron in the single subband regime [23]. This fairly long length scale suggests that screening from the SC is indeed relevant.

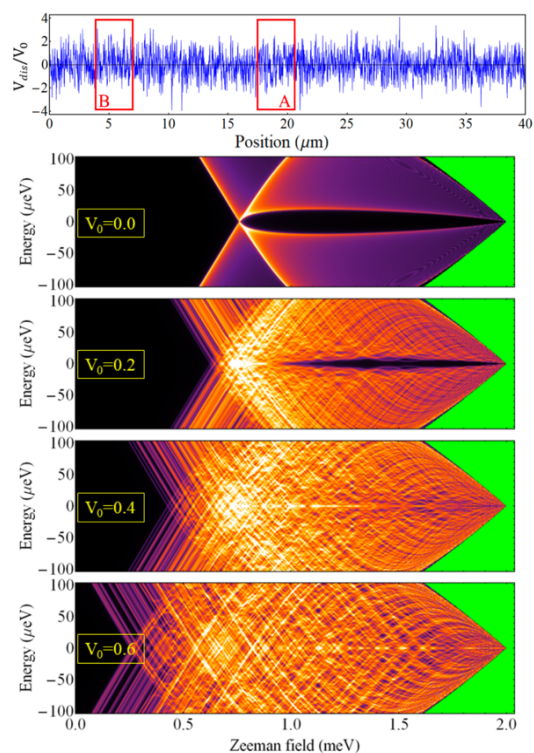

Figure 5. Top: Example of disorder potential landscape for a 40-micron long hybrid. Below: Density of states versus Zeeman energy for increasing disorder amplitudes, $V_o$ (meV). MBS are not included. The topological gap closes for values of $V_o$ several times smaller than the Zeeman gap. From Ref. 52.

It is often assumed that potential variations can be described by a Gaussian distribution with a certain correlation-length and amplitude. Over the past few years there have been numerous papers calculating the effect of disorder on the topological phase in the bulk of 1D hybrid wires. Figure 5 shows exemplary numerical results assuming random disorder [52]. The general phenomenon is that disorder in combination with a magnetic field, breaking time-reversal symmetry, induces an abundant amount of subgap states. For disorder amplitudes of just a fraction of the Zeeman energy (see Fig. 5), the topological gap already completely disappears, in agreement with the rough estimate earlier. (Note that disorder does not affect the zero-field gap due to time-reversal symmetry, illustrating a striking difference between *s*- and *p*-wave SC.) As a result, to obtain a topological bulk phase the amplitude of the disorder potential should remain significantly smaller than ~1 meV. It is unclear whether this has ever been realized in hybrid nanowires. My personal intuition is that with all the possible sources for disorder, current hybrid devices are not clean enough to satisfy the stringent requirements of a bulk topological phase. One really needs to strive towards hybrid devices with everything near perfect on an atomic scale.

**The ends I: barrier disorder.** Figure 5 shows the effect of disorder on the *bulk* topological gap. We now address the issue of disorder in the *barrier* region and how it affects the subgap spectrum at the hybrid ends. If we assume a Fermi energy outside the Zeeman gap and ignore topological effects altogether, then in the presence of disorder our system becomes an SM-SC interface with diffuse scattering in its vicinity. Andreev reflection in combination of many scatterers in a multi-subband nanowire can lead to a zero-bias anomaly, as proposed by several theory papers [53,54,55] as an alternative explanation for our 2012-ZBPs [31]. Indeed, the proposed disorder-ZBPs did look very similar to our Majorana signatures and further experiments were proposed to be able to make a distinction [55]. In any case, the important suggestion was to make the wires cleaner, which became our main focus for the 2012-2016 period, resulting in the ZPBs shown in Figure 2. The ZBP robustness, sticking over a large Zeeman range, as well as the low subgap-conductance background, made us conclude in 2018 [32] that for the cleaner devices we could rule out barrier disorder. The basic argument was that none of the simulations showed a robustness as in Figure 2. In particular, the observation that the ZBPs remained at zero-energy while changing the tunnel barrier from pinch-off to fully open [32] I considered as a strong indication against barrier disorder. Pan and Das Sarma [56] returned to analyzing disorder-ZBPs with extensive simulations in 2022 and re-confirmed that random, short-range disorder can quite generally lead to ZBPs, including peak heights of order $2e^2/h$ [57]. Unfortunately, experimental parameters for disorder, SOI, number of occupied subbands, etc., are not well known, hampering a direct comparison to theory. Question remains if disorder can fully explain the robust ZBPs of Figures 2 and 4. I will return to this question in the next paragraph.

**The ends II: smooth potential.** Our 2012 publication also triggered another class of alternative explanations for ZBPs. Various numerical studies [58-61] pointed out that ZBPs can arise also for trivial reasons when the boundary potentials are not sharp but instead extend over some region, even in the absence of short-length scale disorder. Note that the idealized hard-wall boundary in Figure 1 was implicitly assumed in the nanowire-Majorana proposals. The new numerical models took some artificial landscape with smooth variations in parameters such as electrostatic potential, SOI, induced gap, or Zeeman energy. The general observation was that one can construct parameter landscapes that lead to ZBPs robust in gate voltage and magnetic field, but which do not require disorder, nor a topological phase in the bulk of the 1D hybrid.

Over the years 2012-2016 I was aware of the possible existence of a smoothly varying parameter landscape near the end of the 1D hybrid. Our physical picture of around 2016 is sketched in Figure 6. A hard-wall potential with, for instance, the 3$^{rd}$ subband tuned into the Zeeman gap, is the idealized nanowire-Majorana scenario for observing a topological ZBP at the wire end. Keeping the bulk the same but now for a smooth barrier moves the MBS from the 3$^{rd}$ subband into the bulk decreasing its visibility in a tunneling measurement. In addition, the two lower subbands have small regions with the Fermi energy in the Zeeman gap, resulting in local MBSs. It was thought that such local MBSs would quickly split to finite energies due to their spatial overlap. A ZBP would still result from the topological 3$^{rd}$ subband although with reduced visibility, i.e. a lower peak height. All the reported ZBPs in 2012 only had a ~5% amplitude and this scenario of a weak coupling due to a smooth barrier was thought to be a likely explanation.

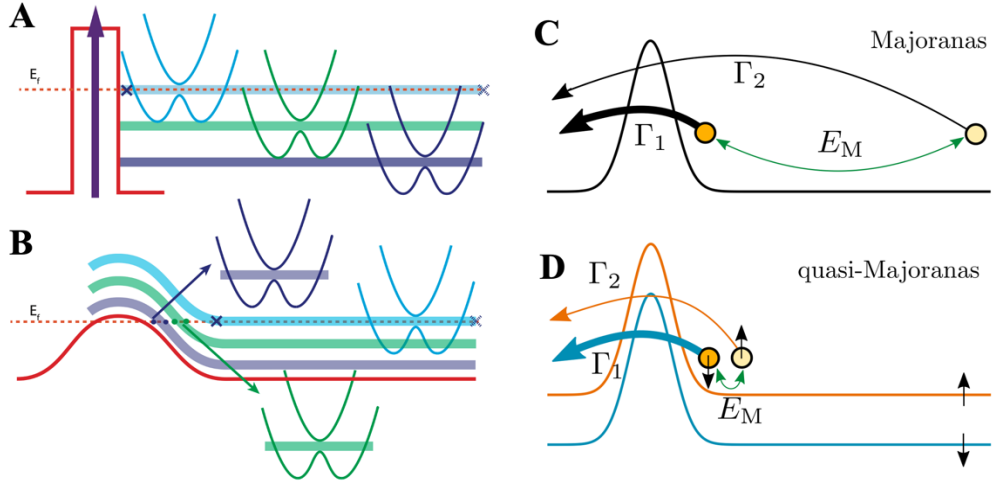

Figure 6. **A.** Hard-wall potential as implicitly assumed in the nanowire-Majorana model with Fermi energy in the Zeeman gap of the 3$^{rd}$ subband. The two crosses schematically indicate MBS locations. **B.** A smooth barrier moves the MBS in the 3$^{rd}$ subband into the bulk, increasing the distance to the outside lead on the left, thereby decreasing its visibility in spectroscopy. The smooth barrier also induces small regions where the Zeeman gap (indicated by purple and green bands) of the 1$^{st}$ and 2$^{nd}$ subband cross the Fermi energy. These crossing points can induce quasi-MBS. From Ref. 62. **C.** Illustration of separated MBS at opposite wire ends such that only one MBS couples to the left lead. **D.** Illustration of spatially-overlapping quasi-MBS from the purple band in **B** that do not split to finite energy and have two opposite spin parts. These two parts can couple very differently to the left lead such that a spectroscopy measurement only probes the outside quasi-MBS. The resulting ZBP is in practise indistinguishable from a topological ZBP. **C** and **D** from Ref. 17.

The quantized Majorana conductance, although retracted later, inspired new theoretical work. Vuik et al. [17] showed that the splitting of spatially overlapping MBSs could be largely suppressed due to an orthogonal spin-structure. (In hindsight, Vuik et al. followed earlier work [58-61] and extensions can be found in [63].) The two local MBSs with opposite spins can differ in the tunnel coupling to the outside lead by orders of magnitude [17]. This can result in a quantized ZBP conductance at $2e^2/h$, in contrast to a $4e^2/h$ plateau if both local MBSs would couple to the lead. The surprising numerical result was that the two local MBSs can be largely decoupled from each other despite considerable spatial overlap. If just one part contributes to the tunneling conductance than its resulting ZBP is indistinguishable from ZBPs due to topological MBSs. The local MBSs were dubbed "quasi-Majoranas" indicating their self-adjoint character on the one hand but lacking a topological character on the other.

The concept of quasi-Majoranas, or quasi-MBSs, was tested further by making the end potential intentionally smoother and check if the resulting conductance quantization of the ZBPs get more robust. This was tested experimentally in Hao Zhang's group in Beijing [64]. They found robust quantization in a range that extended over parameter variations larger than can be assumed for disorder potentials. Moreover, plateaus were only observed within 5% of $2e^2/h$. Conductance values away from $2e^2/h$ smoothly varied with gate and $B$ without showing a plateau. My conclusion is that in-gap, Andreev conductance can be quantized at $2e^2/h$ without invoking a topological bulk phase.

One can identify many reasons that the end of a 1D hybrid by no means can be described by a hard-wall potential. The electrostatic potential is screened differently underneath the SC than in the uncovered sections. This potential variation can be large when the barrier is close to pinch-off while the hybrid has multiple subbands occupied. There often is some distance between the tunnel barrier and the hybrid where SOI, effective $g$-factor and induced gap can vary smoothly. The SC induces strain in the SM and the strain potential varies when crossing the hybrid's end. Moreover, this strain variation may change while cooling down the sample to low temperatures. Finally, the band bending is different at the SM-SC interface compared to SM-vacuum. This difference could be (many) tens of meVs and thus a very significant variation that is not easily screened away.

The resulting sum of all these variations by no means needs to be monotonic. In fact, quite often unintentional quantum dots are found in the barrier region [36], indicating a dip in the potential. Numerical models often cook up an artificial landscape or calculate a self-consistent potential including a subset of the above causes. Unfortunately, we lack a quantitative understanding of the various contributions and in that sense also the hybrid's end is a black box.

It has become apparent that the combination of all possible causes for both disorder as well as smooth parameter variations at the hybrids end lead to ZBPs that have some robustness in gate voltage and magnetic field. It could well be that all the reports on ZBPs in 1D hybrid nanowires just measure end properties that have no relation to the properties in the bulk part of the nanowires. In fact, there is no solid proof against non-topological scenarios' and I personal believe that indeed all reported ZBPs in 1D hybrids have a trivial origin.

Some clean data sets like in Figures 2 and 4, I feel are difficult to explain by short-length scale disorder only. The combination of bulk disorder that fills a possible topological gap completely with subgap states (Figure 5) together with smooth-barrier effects most likely can explain all the reported ZBPs, at least from our group. To be completely clear, in my opinion and in hindsight, our Delft experiments have shown no solid evidence for a 1D topological $p$-wave superconductor.

This personal belief applies in fact to all reported hybrid nanowire materials combinations, which basically means the InAs and InSb SMs and the Nb-based and Al-based SCs. There is maybe one exception where the situation is not clear to me, hybrids where a π-phase difference can be applied between two SCs both proximitizing the same nanowire [65]. Nanowires covered all around with SC, i.e. the full-shell hybrids, also fall in this category since a flux through the wire can effectively induce a π-phase difference between opposite sides of the nanowire [66]. The physics that could result in a topological phase is different from the nanowire-Majorana models and, for instance, could still exist in the case of many subbands occupied; see for a recent theoretical study [67]. Nevertheless, and unfortunately, also these π-phase systems are not protected against ZBPs from disorder or smooth potentials and negative results have been reported as well [68].

**Non-local I: two-end correlations**

The observation that ZBPs occur ubiquitously in 1D hybrids implies that they can also occur coincidentally simultaneously at both wire ends without any causal relation to each other. The occurrence of ZBPs by no means implies that there are just two zero-energy states in the entire hybrid and that these two together belong to a single fermionic state. Boundary effects and disorder can result in multiple zero-energy states and thus the ones that are probed at the wire ends could have no relation to each other. This seems to be the most likely scenario of all reported ZBPs (maybe again an exception for the full-shell hybrids). Our extensive parameter study [43] as well as the Microsoft paper [23] shows an abundance of ZBPs with only a small fraction occurring simultaneously. This small fraction could reflect statistical coincidences.

"Smoking gun" evidence for topological MBSs have been proposed [69] if two ZBPs measured at opposite ends both show splitting-oscillations due to a finite length of the hybrid and therefore an overlap of the tails of their wavefunctions. If the splitting-oscillations measured from both ends have the same pattern, meaning that amplitudes and zero-crossings are highly correlated, then this could form significant evidence. In particular, if one of the barrier-gates can influence this pattern on its own end but also non-locally at the other end. Such a very basic identifier for topological MBSs have not been reported, not in 1D hybrids and not in 2DEG hybrids. I believe that splitting-correlations in wires much longer than the SC coherence length could still be important evidence for topological MBS. Whether this is full-proof, smoking gun evidence needs detailed scrutiny, like a simultaneous measurement of a bulk gap reopening, as we discuss next.

**Non-local II: non-local transport**
Measuring local transport, even simultaneously from both ends, therefore do not provide much information of the bulk properties of the 1D hybrid (leaving splitting-correlations aside). An interesting technique was proposed [70] to measure non-local properties by applying a voltage on one end and measure the current at the other end. This non-local current can only be non-zero if there exist states, e.g. subgap states that connect the two ends. In long wires, such states with the lowest energies are at the induced gap. If this induced gap decreases with $B$, closes and reopens, then this could be interpreted as evidence for a topological phase transition.

In finite-length hybrids there could still be alternative interpretations but for very long hybrids gap closing and reopening likely indicates a quantum phase transition. Non-local transport has been reported, providing interesting new information [71,23,48]. In 1D hybrids no evidence was found for a phase transition. Gap closing and reopening was found in rare cases, but without simultaneous detection of correlated ZBPs. The gap reopening could well be due to finite size effects [48]. Microsoft developed a topological gap protocol [72] and presented experiments satisfying this protocol [23]. The protocol, however, contains hidden assumptions that may not correspond to the real device situation. Disorder-induced ABSs and smooth barrier effects together can give false-positives on the topological gap protocol [73,74,75]. It remains to be seen whether passing the topological gap protocol indeed implies a topological phase in the bulk of the Microsoft hybrids. There is still an ongoing discussion both at an experimental level as well as about the theoretical interpretation [73-76]. My perspective is that the data presented in the MSFT paper shows many features due to disorder and I hope that future results show much improved data quality.

**Support from Microsoft**
My presence at the Station Q meeting in 2012 led to financial support from Microsoft for academic research in Delft. The support was organized as Research Project Descriptions, approved, and administered by the Dutch Science Foundation (FOM/NWO) and TU Delft. Besides a running budget, the funding enabled purchasing three dilution refrigerators (from Leiden Cryogenics) with measurement setups for a total of roughly 3 M€. In addition, I could hire 20 PhD students and 10 postdoctoral researchers in the period 2012-2022. The Microsoft support funded my research almost completely during this period. All the output of this research has been published in journals and always with open-access (preprint) versions on arXiv.

In November 2016 my employment transitioned from TU Delft to Microsoft. I kept my professor rights at TU Delft allowing me to graduate (PhD) students. All other academic responsibilities stopped. During my Microsoft employment academic research with (PhD) students continued as before including the open-access publications. Besides this academic research I was involved with other Microsoft business which is proprietary and confidential. I left Microsoft March 2022 and re-focused on academic research at TU Delft and that continues as of today.

The research on 1D hybrids over this entire period was administered entirely by TU Delft. All output is available open-access on arXiv.org, as well as through the repository for PhD thesis's at TU Delft:
https://repository.tudelft.nl/islandora/search/Kouwenhoven?f%5B0%5D=mods_genre_s%3A%22doctoral%5C%20thesis%22.

**Summary and conclusions.** The enthusiasm inspired by the physics of Majorana particles, and the promise of error-protected topological quantum computing has generated an enormous activity on SC-SM hybrids. We have seen great progress with innovative new devices that include gate-controlled transmon qubits (i.e. gatemons), Andreev spin qubits, Cooperpair splitters and Kitaev chains [77]. We have learned extensively on many new aspects of mesoscopic superconductivity down to the microscopic level of individual quantum states, that include ABS with their anomalous charge and spin, as well as all kinds of zero-energy states with either a trivial or topological character [1,16].

We have learned that zero-energy states are abundant in 1D hybrids. Breaking time-reversal symmetry by applying a magnetic field opens a Pandora box for subgap states. The signals of these states in local and non-local transport can mimic almost perfectly the predicted signatures for MBS in the nanowire-Majorana model, making it very hard to distinguish topological from trivial states. I cannot exclude that InAs/Sb-based hybrids will ever fulfill the requirements for the nanowire-Majorana approach, but I doubt it. I think it is time to invest in developing new materials with improved parameters for spin-orbit strength and $g$-factor and turn these new hybrid materials into devices with much lower levels of impurities. PbTe nanowires are an interesting candidate [78]. Or completely different platforms like the 2D van der Waals materials [79]. Hand in hand with the experiments, there is a need for device simulations that include issues like grain structures in the SC, band bending at the hybrid interface as well as at the uncovered surface [80], induced strain from the SC-SM lattice mismatch, including inhomogeneous strain from the SC grain structure. I emphasize these three issues since they involve energy scales much larger than the topological gap. The theory community is ignoring these issues, probably since they are difficult to model and moreover, they are not prestigious subjects for quantum theorists.

Until at least 2017 we felt that only MBS models were consistent with our experiments, despite that we knew we had no evidence for a topological phase. Other possible causes like disorder-induced ABS including level repulsion [33], or the Kondo effect [81], we felt could all be excluded. This was the status of our mindset in 2017, for instance, as expressed in [62]. The possibility of a smooth barrier for inducing quantized ZBPs without a topological phase came to me with the work of Vuik et al. [17]. Disorder has always been a concern, but I didn't think we needed the extreme device quality similar to the 5/2 fractional quantum Hall effect but now in combination with superconductivity. The devastating effects from disorder came to me only with the simulations from post 2020 by Das Sarma and colleagues from Maryland, see Figure 5 as an example. The question that comes up is "Why, despite early theoretical warnings, was the importance of a smooth barrier and disorder largely ignored for the first 5-6 years?"

A scholar.google search on "Majorana nanowire" in the 2012-2018 period returns ~3500 papers, that is every day ~2 papers. I probably clicked on most of them but have not read them all carefully. I selected papers for a full reading, and these included for instance [59,60]. Re-reading these papers now, I clearly see predictions for smooth-barrier ZBPs. At the time, I also read several disorder papers [54,55] but it seemed to me these papers assumed unrealistically large amounts of disorder. I discussed in person with many colleagues in the field every aspect of possible criticism and ideas for new experiments, including at more than 50 international meetings. These interactions did not shout out for smooth barriers or disorder as the origin for our ZBPs. Vuik et al.'s work on smooth barriers [17] had a clarity that for me connected the dots. Also, Das Sarma's simulations [56,57] had graphs that were "in-your-face' and by no means could be ignored. Since then, smooth-barriers and disorder are on my, and everybody else's, radar.

I must note here that discussions I had in the past two years have retrospectively cast a more critical light on the 2012-2016 period than was present during that time. The early heightened attention created a strong dominant voice roughly saying, '*now that we have Majoranas, a topological qubit is next*'. There was no room for critical voices saying, '*I don't think you have Majoranas*'. The collective excitement pushed the focus and the funding towards the next phase of qubit development, despite the weak fundamental basis for Majoranas. I guess optimism became opportunism.

The enthusiasm for Majorana research, thus came with some serious blind spots. We knew that the long hybrids were a black box in many aspects, but I guess we hoped for the best of it. Maybe slowly but nevertheless we learned, faced and identified many of the earlier unknowns. And the good news is that we now have bright lights on these blind spots, and it is time for renewed optimism, as discussed in the **Future** section.

**Lessons learned.** Sometimes condensed matter experiments provide clear observations, like quantum Hall plateaus. Sometimes results have an unambiguous interpretation, like Rabi-oscillations demonstrating a qubit. But there are also research fields that are complicated and 'messy', where results are not easy to reproduce and where a common understanding is lacking. Research on MBS and topological SC is on this side of the spectrum. $Sr_2RuO_4$ is an illustrative example. It was considered as a candidate for being a naturally occurring p-wave SC supported by ample reported evidence taken over 3 decades [82]. However, more precise measurements showed the opposite, and it is no longer believed to be a topological SC [82]. Our field of MBS has seen a similar evolution. To some extent, we must accept that a research topic turns out to be complicated and that there can be phases of hope and promise followed by disappointment. I personally have tended to hold on to the phases of hope and promise and ignore counter signs. 'To keep faith against all odds' helps to solve a big challenge. This, however, should come hand in hand with experimental due diligence and carefully dealing with facts and data. Here, we obviously came short with the two Nature retractions.

I know that several of my graduate students felt (peer) pressure to produce impactful results. The hype, the (media) attention, the luxurious funding, the field primarily publishing in Science and Nature, it all adds up towards high expectations. I am not aware if these expectations ever led directly to actions like cutting on due diligence. I always felt seriousness and dedication from all group members. But the circumstances were extraordinary and most likely it did affect my group.

To keep an appropriate balance between internal optimism versus external critique several conditions need to be satisfied, i.e. (a) transparency and availability of data; (b) representative presentation of available results; (c) reproducibility of experiments. I honestly think we have satisfied on (a) and (c), although one can always argue to what extend unpublished data should be disclosed. (b) is more complicated. Our working method has been to demonstrate proof-of-concepts. A discovery or demonstration of a prediction just needs one example to show that it is possible in principle. Journals are satisfied if two examples can be shown, to satisfy the reproducibility requirement. Since we spent a full decade on MBS, our collective research and output provides a comprehensive overview of the various experiments, including their connections and their reproducibility. Taken individual publications, then the positive promises sometimes overshadowed the complications. One thing that should have been emphasized much more is that many of the published figures were taken on our best performing devices, our so-called 'hero devices'. An occasional hero-device publication is fine, in my opinion, but if a research field is simply a sequence of hero-device reports then a bubble-problem can arise. The MBS research has had an excessive emphasis on proof-of-concepts from hero devices. This working method is fertile ground for a culture of confirmation bias; there is a prediction and that needs to be proven. (In essence a similar view on the field was described in a recent Comment in Nature by Sergey Frolov [83].) I personally should have recognized this earlier and acted on it. The good thing of the retracted Nature publications is that this process came to a stop. The downside of course is that it came to a complete stop. There were no talks on nanowire-Majorana progress at the 2024 APS meeting in Minneapolis. Currently, I only know of ongoing nanowire-Majorana activities in Beijing [59] and at Microsoft [22]. But I do believe we can pull up again, as described next in the Future section.

**Future.** The future is a lot brighter compared to a decade ago. Materials, fabrication, simulations, measurement techniques as well as our understanding, have all improved significantly. I see four interesting future directions.

First, there are two ongoing efforts on the InAs-Al platform. Microsoft is pursuing their 2DEG approach. From the 100+ authors on their papers I have the impression that they are brute forcing all the challenges and squeeze out whatever is possible. Whether that is enough for a qubit and a braiding demonstration is unclear. I do hope for a disclosure of their tricks and hacks so that the learnings can be used in academia as well. The other InAs-Al direction that needs clarity are the full-shell nanowires [66]. The full-shell wire results have been criticized including an editorial expression of concern [84] creating confusion with respect to the interesting findings in [66]. This line of research needs a clear conclusion.

Next, what I find promising are alternative materials with stronger SOI and thereby more resilient against developing subgap states in a magnetic field due to disorder. A larger parent SC gap has the potential to increase the topological gap. Strong-SOI/large-SC gap hybrids should also have an 'ease of fabrication' such that the pristine materials remain clean. It would be helpful to have extended simulations on materials and devices to assist in an efficient search for good candidates.

Finally, our group in Delft recently found a way around the Pandora box of disorder and smooth potentials. We developed fabrication technology [42] for realizing the original proposal by Kitaev [85]. In a chain of quantum dots separated by short hybrids (~200 nm), tunneling and cross-Andreev reflection can be tuned towards a 'sweet spot' with MBS [19]. Even a minimal two-site chain yields two MBS localized on the quantum dots, separated by the hybrid section [18]. In this case, the ZBPs are associated with "poor man's Majoranas" [86]. We now have a multi-group collaboration in Delft on Kitaev chains that recently realized a three-site device [87] that clearly shows gap-closing and re-opening concurrently at the point where MBS appear on the outside quantum dots. The "poor man's" adjective was dubbed as a reminder that short chains have a 'short bulk' yielding a gap in a discrete energy spectrum (in contrast to continuous energy bands) and that the MBS protection is only partial and not topological [86]. For long chains it has been shown theoretically [19] that a topological $p$-wave phase should develop in the bulk section.

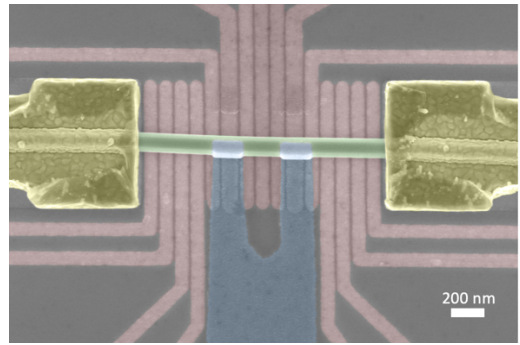

Figure 7. Picture of a 3-site Kitaev chain with two hybrid sections (blue) connected by a flux-loop. A total of 15 gate electrodes (purple) allows tuning to the sweet spot and perform parity readout with charge sensors defined by the outermost gates. Picture from 'work in progress'.

Disorder in a Kitaev chain just implies retuning the gate values for compensation. Our smart-wall fabrications yield very stable devices so that the sweet spot is stable for weeks. There is a clear path towards longer arrays with a topological phase in the bulk. There is also a clear cost of having to operate more gates, but we have learned already that tuning follows a strict protocol that is suitable for a machine learning tuning procedure [88]. Our focus in the coming years will be on short Kitaev chains, build parity qubits and perform braiding with 'poor man's' Majorana bound states. In any case, this direction of Kitaev chains is not based on black boxes and its success does not depend on hero devices and proof-of-principle experiments.

    A decade of research on nanowire-Majoranas received significant attention when it was upcoming [2,14,15] as well as when it suffered from paper retractions and heavy criticism [83]. My purpose for this Perspective is to provide some clarity in the risen confusion and to share my experiences during this decade.

**Research Volume (2010-2023): 8 PhD thesis's at repository.tudelft.nl with focus on realizing the Nanowire-Majorana model**

I. van Weperen (2014), Quantum Transport in Indium Antimonide Nanowires: Investigating building blocks for Majorana devices
K. Zuo and V. Mourik (2016) Signatures of Majorana Fermions in Hybrid Superconductor-Semiconductor Nanowire Devices
D.J. van Woerkom (2017) Semiconductor Nanowire Josephson Junctions: In the search for the Majorana
Önder Gül (2017) Ballistic Majorana nanowire devices
M.W.A. de Moor (2019) Quantum transport in nanowire networks
J.D.S. Bommer (2021) Zero-energy states in Majorana nanowire devices
D. Xu (2022) Quantum Properties in Hybrid Nanowire Devices
N. van Loo (2023) Shadow-wall lithography as a novel approach to Majorana devices

**Data repositories including analysis and corrections can be found at:**

M.W.A. de Moor et al., *Electric field tunable superconductor-semiconductor coupling in Majorana nanowires*, New J.Phys. **20**, 103049 (2018): https://zenodo.org/records/7679180
Ö. Gül et al., *Hard Superconducting Gap in InSb Nanowires*, NanoLett. **17**, 2690 2017): https://zenodo.org/records/7729730
J.D.S. Bommer et al., *Spin-Orbit Protection of Induced Superconductivity in Majorana Nanowires*, Phys.Rev.Lett. **122**, 187702 (2019): https://zenodo.org/records/7671990
H. Zhang et al., *Ballistic superconductivity in semiconducting nanowires*, Nat.Commun. **8**, 16025 (2017): https://zenodo.org/records/6851435
E. Fadaly et al., *Observation of Conductance Quantization in InSb Nanowire Networks*, NanoLett. **17**, 6511 (2017): https://zenodo.org/doi/10.5281/zenodo.4989951
Ö. Gül et al.,*Ballistic Majorana nanowire devices*, Nature Nanotechnology **13**, 192 (2018): https://zenodo.org/doi/10.5281/zenodo.4721356

**Acknowledgement**

I thank all team members during the 2010-2020 decade for their valuable contributions. I thank our 2020-2024 team members for keeping their focus on the physics and successfully create new enthusiasm with the Kitaev chains. For their comments on this Perspective, I thank Ramon Aguado, Silvano De Franceschi, Önder Gül, Nick van Loo, Daniel Loss, Charles Marcus, Greg Mazur, Elsa Prada, Jay Sau, Di Xu, Guanzhong Wang, Michael Wimmer, and Hao Zhang. I acknowledge Sergey Frolov and Vincent Mourik for their scientific criticism.

**References.** This list is limited to papers most relevant to this Perspective. Ref. 1 provides a comprehensive list of the field.